\documentclass{iopart}
\usepackage[dvipdfm]{graphicx}
\newcommand{\be}{\begin{equation}}
\newcommand{\ee}{\end{equation}}

\begin{document}
\title{A new correlator in quantum spin chains}
\author{J.P. Keating, F. Mezzadri and M. Novaes}
\address{School of Mathematics, University of Bristol, Bristol BS8
1TW, UK}

\begin{abstract}
We propose a new correlator in one-dimensional quantum spin chains,
the $s-$Emptiness Formation Probability ($s-$EFP). This is a natural
generalization of the Emptiness Formation Probability (EFP), which
is the probability that the first $n$ spins of the chain are all
aligned downwards. In the $s-$EFP we let the spins in question be
separated by $s$ sites. The usual EFP corresponds to the special
case when $s=1$, and taking $s>1$ allows us to quantify non-local
correlations. We express the $s-$EFP for the anisotropic XY model in
a transverse magnetic field, a system with both critical and
non-critical regimes, in terms of a Toeplitz determinant. For the
isotropic XY model we find that the magnetic field induces an
interesting length scale.
\end{abstract}


\section{Introduction}

The Emptiness Formation Probability (EFP) is defined for a
one-dimensional spin-$1/2$ chain to be the probability of formation
of a ferromagnetic string, i.e. the probability that the first $n$
spins of the chain are all aligned downwards, \be P(n)=Z^{-1}{\rm
Tr}\{\rme^{-H/k_BT}F\},\ee where $H$ is the Hamiltonian of the
system, $Z={\rm Tr}\{\rme^{-H/k_BT}\}$ is the partition function and
\be F=\prod_{j=0}^n\frac{1-\sigma_j^z}{2}.\ee Here $\sigma^z_j$ is
the usual Pauli spin matrix and $j$ labels the sites on the chain.
At zero temperature the EFP is given by \be P(n)=\langle
0|F|0\rangle,\ee where $|0\rangle$ denotes the ground state of the
system. This fundamental correlator has been the subject of a good
deal of recent interest
\cite{pla190vek1994,npb647aga2002,kitanine,pla312vek2003,jpsj70ms2001,jpa38ff2005}.
It usually cannot be calculated exactly, but a number of results
have been obtained regarding its asymptotic behaviour as
$n\to\infty$.

In this Letter we propose a new correlator in one-dimensional
quantum spin chains: the $s$-Emptiness Formation Probability
($s-$EFP). This is a natural generalization of the usual EFP. What
we consider is the probability of finding $n$ equally spaced sites
aligned downward, for a given spacing $s$. Specifically, the $s-$EFP
is defined as \be\label{sefp} P_s(n)=\langle
0|\prod_{j=0}^n\frac{1-\sigma_{sj}^z}{2}|0\rangle.\ee Notice that we
take the Pauli matrices at the first $n$ sites whose labels are
multiples of $s$. The EFP thus corresponds to the particular case
when $s=1$. Since (\ref{sefp}) is not restricted to adjacent spins
the $s-$EFP contains non-local information that is not available
through the EFP. For the anisotropic XY model we shall derive its
asymptotic form as $n\to\infty$ using the theory of Toeplitz
determinants. For simplicity we take the temperature to be zero, but
an extension to finite temperature is straightforward.

\section{EFP and $s-$EFP for the XY model}

Abanov and Franchini have recently considered the EFP for the
anisotropic XY model in a transverse magnetic field $h$
\cite{jpa38ff2005}. They used the theory of Toeplitz determinants
and extensions of what is generally known as the Fisher-Hartwig
conjecture \cite{fisher,widom} to obtain the asymptotic
($n\to\infty$) behaviour of $P(n)$. The XY Hamiltonian is
\be\label{xy}
H=\sum_{j=0}^{N-1}\left(\frac{1+\gamma}{2}\sigma_j^x\sigma_{j+1}^x+
\frac{1-\gamma}{2}\sigma_j^y\sigma_{j+1}^y\right)-h\sum_{j=0}^{N-1}\sigma_j^z,\ee
and it is critical along the lines $\gamma=0,|h|<1$ and $h=\pm 1$,
i.e. quantum correlations decay algebraically when the parameters
are on these lines and exponentially when they lie away from it
\cite{book,pra3eb1971}. The EFP can in this case be written as
\be\label{deter} P(n)=|{\rm det}S_n|,\ee where $S_n$ is an $n\times
n$ Toeplitz matrix whose elements are \be
\left(S_n\right)_{jk}=\frac{1}{2\pi}\int_0^{2\pi}\rme^{\rmi(j-k)\theta}\sigma(\theta)d\theta.\ee
The function $\sigma(\theta)$ is called the symbol of the matrix
$S_n$, and is given by \be\label{sigma}
\sigma(\theta)=\frac{1}{2}+\frac{1}{2}\frac{\cos\theta-h+\rmi\gamma\sin\theta}
{\sqrt{(\cos\theta-h)^2+\gamma^2\sin^2\theta}}.\ee

Abanov and Franchini obtained the asymptotic form of the determinant
(\ref{deter}) in all regions of the $\gamma-h$ phase diagram. Their
results are as follows \cite{jpa38ff2005}: for $1>h\neq-1$,
$P(n)\sim E\rme^{-n\beta}$; for $h>1$ the previous result is
multiplied by $[1+(-1)^nA]$; on the critical lines $h=\pm1$,
$P(n)\sim En^{-1/16}[1+(-h)^nA/\sqrt{n}]\rme^{-n\beta}$; and on the
critical line $\gamma=0,|h|<1$, $P(n)\sim
En^{-1/4}\rme^{-n^2\alpha}$, i.e. the decay is Gaussian. Explicit
formulae were obtained for $E$, $A$, $\alpha$ and $\beta$ as
functions of $\gamma$ and $h$. The latter, for example, is given by
\be
\beta=-\frac{1}{2\pi}\int_0^{2\pi}\rmd\theta\log|\sigma(\theta)|.\ee

For this system the $s-$EFP corresponds simply to \be P_s(n)=|{\rm
det}S_{n}(s)|,\ee where $S_{n}(s)$ is obtained from $S_n$ by
removing the rows and columns whose labels are not multiples of $s$.
Its matrix elements are \be
\left(S_{n}(s)\right)_{jk}=\frac{1}{2\pi}\int_0^{2\pi}\rme^{\rmi
s(j-k)\theta}\sigma(\theta)\rmd\theta.\ee This is in Toeplitz form,
but $\sigma(\theta)$ is no longer the symbol. To determine the
symbol we must find the function $\sigma_s$ that satisfies
\be\label{gtoh} \int_0^{2\pi}\sigma(\alpha)\rme^{-\rmi
sn\alpha}\rmd\alpha=\int_0^{2\pi}\sigma_s(\alpha)\rme^{-\rmi
n\alpha}\rmd\alpha.\ee Multiplying (\ref{gtoh}) by $\rme^{in\theta}$
with $0\leq\theta<2\pi$, summing over $n$ and using the Poisson
summation formula, we arrive at \be\label{aver}
\sigma_s(\theta)=\frac{1}{s}\sum_{n=0}^{s-1}\sigma\left(\frac{\theta}{s}+\frac{2n\pi}{s}\right).\ee
The value of $\sigma_s$ at the point $\theta$ is therefore obtained
as the average value of the original symbol $\sigma$ over the $s$
vertices of a regular polygon. Hence the $s-$Emptiness Formation
Probability of the XY model (\ref{xy}) can also be computed as a
Toeplitz determinant.

Given the parameters $\gamma$ and $h$, the numerical calculation of
the average (\ref{aver}) is a trivial task. Exact analytical
results, on the other hand, are only available for particular cases.
One of these cases is the isotropic XY model, which we analyze in
the next section.

\section{The isotropic XY model}

As a simple application, let us consider the isotropic XY model, for
which $\gamma=0$. In the non-critical regime $|h|>1$ the ground
state is always a ferromagnet, i.e. the spins are all aligned up if
$h>1$, and all aligned down if $h<-1$. Therefore $P(n)$ vanishes in
the first case and is unity in the second. Since the chain is
translation-invariant, $P_s(n)$ is the same as $P(n)$ for any choice
of $s$.

The critical regime $|h|<1$ is of course much more interesting. The
symbol (\ref{sigma}) reduces to \be \sigma(\theta)=\cases{1\;
 {\rm if }\;-k\leq\theta<k,\\0 \;{\rm otherwise},} \ee where
$k$ is related to the magnetic field by \be h=\cos k.\ee Applying
Widom's theorem \cite{widom} to the resulting Toeplitz determinant,
one concludes that as $n\to\infty$ the EFP behaves like $P(n)\sim
En^{-1/4}e^{n^2\alpha}$, where $E$ and $\alpha$ are explicitly
determined constants. In order to obtain the $s-$EFP we must compute
the average (\ref{aver}), which in this case can be done explicitly.
The result is a piecewise constant and even function, with jumps at
the critical points $\pm\theta^\ast$ given by \be
[0,\pi)\ni\theta^\ast={\rm min}\{\alpha,2\pi-\alpha\}, \qquad
\alpha=sk\;{\rm mod }\;2\pi,\ee and values \be\label{aga}
\sigma_s(\theta)=\frac{1}{s}\left(1+2\left[
\frac{sk}{2\pi}\right]\right)+\cases{0\; {\rm if }\;
|\theta|<\theta^\ast,\\ \epsilon/s\; {\rm otherwise},}\ee where the
brackets $[\cdot]$ denote the integer part and \be \epsilon={\rm
sign}\{\alpha-\pi\}.\ee

If $s<\pi/k$ we have $\sigma_s(0)=1/s$ and $\sigma_s(\pi)=0$. In
this case we get simply $P_s(n)=s^{-n}P(n)$, so the $s-$EFP decays
much faster than the EFP, combining the original Gaussian decay with
an additional exponential term that depends on the spacing. On the
other hand, if $s>\pi/k$ then the function $\sigma_s(\theta)$ is
never zero, and instead of applying Widom's theorem we should apply
the Fisher-Hartwig conjecture \cite{fisher}. In that case the decay
is no longer Gaussian, but becomes an exponential with a power-law
prefactor, \be P_s(n)\sim En^{-\xi}e^{-\beta n}.\ee The constant $E$
can be obtained from the Fisher-Hartwig formula and the other
quantities are \be\label{beta} \beta(s)=-\frac{1}{\pi}[\theta^\ast
\ln(\sigma_s(0))+(\pi-\theta^\ast)\ln(\sigma_s(\pi))]\ee and \be
\xi=\frac{1}{4\pi^2}\ln^2(\sigma_s(\pi)/\sigma_s(0)).\ee

Interestingly, the decay of $P_s(n)$ therefore shows a clear
transition at the magnetic-field-dependent value $s=\pi/k$, changing
from Gaussian to exponential. In particular, for large spacings we
see that both $\sigma_s(0)$ and $\sigma_s(\pi)$ tend to $k/\pi$.
Hence in this limit there is no power-law prefactor and the exponent
saturates: \be \lim_{s\to\infty}\beta(s)=\ln(\pi/k).\ee

It is important to observe that expression (\ref{beta}) for the
exponent $\beta(s)$ is continuous when we consider $s$ as a real
number, despite the discontinuities that appear in (\ref{aga}). The
function $\sigma_s(0)$ is discontinuous whenever $sk=2n\pi$, but at
those points $\theta^\ast$ vanishes and hence $\beta(s)$ remains
unaffected. The same is true for the jumps in the function
$\sigma_s(\pi)$, which occur at $sk=(2n+1)\pi$ (due to the variable
$\epsilon$), because then we have $\theta^\ast=\pi$. The derivative
$\beta'(s)$, on the other hand, is discontinuous: at the special
points $sk=n\pi$ the function $\beta(s)$ has a local minimum with a
cusp form. Remarkably, for a fixed $k$ its value at the minimum is
independent of $n$, and is actually equal to its $s\to\infty$ limit,
\be \beta(n\pi/k)=\ln(\pi/k).\ee

\begin{figure}[t]
\includegraphics[scale=0.33,angle=-90,bb=50 -180 550 405]{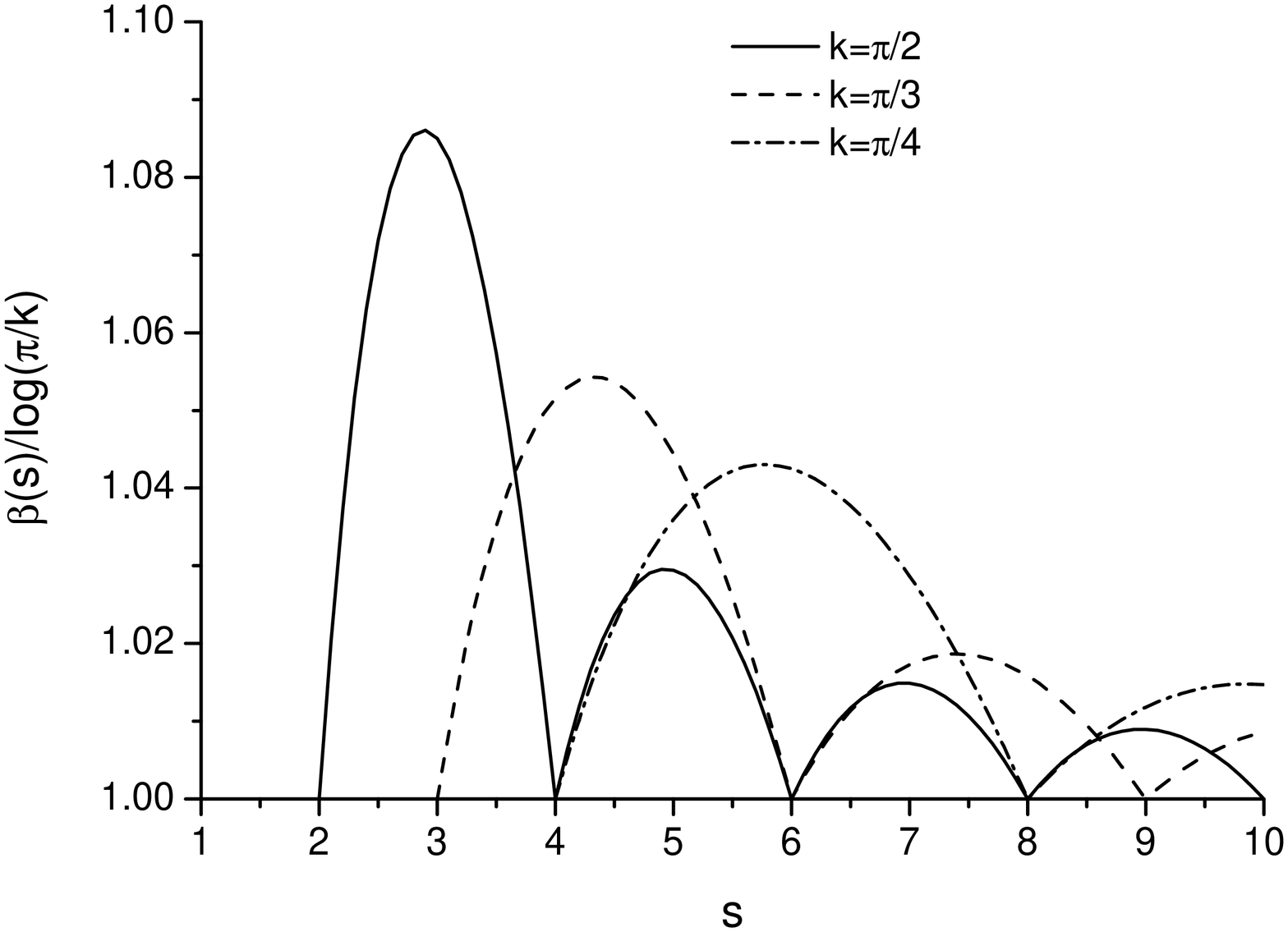}
\caption{The renormalized exponent $\beta(s)/\log(\pi/k)$, defined
in (\ref{beta}), for some values of the magnetic field $h=\cos k$.
We see that for $k=\pi/\ell$ there exists a typical length scale for
the spacing $s$: the exponent is minimal for $s=n\ell$, and its
value is equal to its asymptotic value. Notice that the exponent
$\beta(s)$ is not defined for $s<\ell$ because in this regime the
decay is Gaussian. The point $s=\ell$ marks a crossover in the
behaviour of $P_s(n)$.}
\end{figure}

In Fig.1 we plot the function $\beta(s)$ for different values of the
magnetic field $h=\cos k$. Since, of course, only integer values of
$s$ may be realized in the actual chain, we pick values of the form
$k=\pi/\ell$ with an integer $\ell$. We see that this leads to the
appearance of the above mentioned new length scale: $\beta(s)$
attains its minimal value whenever $s$ is a multiple of $\ell$. It
is worth remarking that Toeplitz determinants have also been used to
describe quantum entanglement
\cite{jsp116bqj2004,cmp252jpk2004,jpa38ari2005}, and that a
generalization of the usual entanglement geometry, similar to the
one proposed here, involving spins that are $s$ sites apart,
exhibits behaviour like that of the $s-$EFP \cite{preprint}.

\section{The line $\gamma^2+h^2=1$}

In the phase diagram of the XY model the line $\gamma^2+h^2=1$,
which Franchini and Abanov call $\Gamma_E$, is special. Along it the
ground state is completely disentangled \cite{line}, i.e. it is
given by \be |0\rangle_{\Gamma_E}=\prod_j
[\cos(\vartheta/2)|\uparrow\rangle_j+(-1)^j\sin(\vartheta/2)|\downarrow\rangle_j],\ee
where $|\uparrow\rangle_j$ and $|\downarrow\rangle_j$ denote spin-up
and spin-down states respectively at the site $j$, and the parameter
$\vartheta$ is such that
\be\cos\vartheta=\sqrt{\frac{1-\gamma}{1+\gamma}}.\ee Let us
consider, for simplicity of notation, only the case $h>0$. Then the
symbol (\ref{sigma}) is given by \be
\sigma_{\Gamma_E}(\theta)=\frac{1}{2}\left(1+\frac{z-\cos\vartheta}{1-z\cos\vartheta}\right),\ee
where $z=\rme^{i\theta}$. The average (\ref{aver}) becomes \be
\left(\sigma_{\Gamma_E}\right)_s(\theta)=\frac{1-\cos\vartheta}{2}+
\frac{\sin^2\vartheta}{2}\frac{z\cos^{s-1}\vartheta}{1-z\cos^s\vartheta},\ee
and its modulus is \be
\left|\left(\sigma_{\Gamma_E}\right)_s(\theta)\right|=\frac{1-\cos\vartheta}{2}
\left|\frac{1+z\cos^{s-1}\vartheta}{1-z\cos^s\vartheta}\right|.\ee

As expected, the $s-$Emptiness Formation Probability in this case is
independent of $s$, because the exponent $\beta$ is given by \be
\beta=-\log\left(\frac{1-\cos\vartheta}{2}\right),\ee and we thus
have \be
P_s(n)=\left(\frac{1-\cos\vartheta}{2}\right)^n=\sin^{2n}(\vartheta/2),\ee
which is in fact an exact result.

\ack{MN wishes to thank CAPES for financial support. JPK is
supported by an EPSRC Senior Research Fellowship.}


\section*{References}

\begin{thebibliography}{99}

\bibitem{pla190vek1994} Korepin V E, Izergin A G, Essler F H L and Uglov D B 1994
Correlation functions of the spin-$\frac{1}{2}$ XXX antiferromagnet
{\it Phys. Lett. A} {\bf 190} 182

\bibitem{npb647aga2002} Abanov A G and Korepin V E 2002 On the probability of ferromagnetic strings in
antiferromagnetic spin chains {\it Nucl. Phys. B} {\bf 647} 565

\bibitem{kitanine} Kitanine N, Maillet J M, Slavnov N A and Terras V 2002 Correlation functions of the
XXZ spin-$\frac{1}{2}$ Heisenberg chain at free fermion point from
their multiple integral representations {\it J. Nucl. Phys. B} {\bf
642} 433\nonum Kitanine N, Maillet J M, Slavnov N A and Terras V
2002 Emptiness formation probability of the XXZ spin-$\frac{1}{2}$
Heisenberg chain at $\Delta=\frac{1}{2}$ {\it J. Phys. A} {\bf 35}
L385\nonum Kitanine N, Maillet J M, Slavnov N A and Terras V 2002
Large distance asymptotic behaviour of the emptiness formation
probability of the XXZ spin-$\frac{1}{2}$ Heisenberg chain {\it J.
Phys. A} {\bf 35} L753

\bibitem{pla312vek2003} Korepin V E, Lukyanov S, Nishiyama Y and Shiroishi M 2003 Asymptotic behaviour
of the emptiness formation probability in the critical phase of the
XXZ spin chain {\it Phys. Lett. A} {\bf 312} 21

\bibitem{jpsj70ms2001} Shiroishi M, Takahashi M and Nishiyama Y 2001
Emptiness formation probability for the one-dimensional isotropic XY
model {\it J. Phys. Soc. Japan} {\bf 70} 3535

\bibitem{jpa38ff2005} Abanov A G and Franchini F 2003 Emptiness formation probability for the anisotropic
XY chain in a magnetic field {\it Phys. Lett. A} {\bf 316} 342\nonum
Franchini F and Abanov A G 2005 Asymptotics of Toeplitz determinants
and the emptiness formation probability for the XY spin chain {\it
J. Phys. A} {\bf 38} 5069

\bibitem{fisher} Fisher M E and Hartwig R E 1968 Toeplitz determinants, some applications, theorems and
conjectures {\it Adv. Chem. Phys.} {\bf 15} 333\nonum Widom H 1973
Toeplitz determinants with singular generating functions {\it Am. J.
Math.} {\bf 95} 333\nonum Basor E L 1978 Asymptotic formulas for
Toeplitz determinants {\it Trans. Amer. Math. Soc.} {\bf 239} 33
\nonum Basor E L and Tracy C A 1991 The Fisher-Hartwig conjecture
and generalizations {\it Physica A} {\bf 177} 167\nonum Basor E L
and Morrison K E 1994 The Fisher-Hartwig conjecture and Toeplitz
eigenvalues {\it Lin. Alg. App.} {\bf 202} 129\nonum Ehrhardt T and
Silbermann B 1997 Toeplitz determinants with one Fisher-Hartwig
singularity {\it J. Funct. Anal.} {\bf 148} 229

\bibitem{widom} Widom H 1971 The strong Szeg{\"o} limit theorem for circular arcs
{\it Ind. Univ. Math. J.} {\bf 21} 277

\bibitem{book} Sachdev S 2000 {\it Quantum Phase Transitions}
(Cambridge: Cambridge Univ. Press)

\bibitem{pra3eb1971} Baruch E, McCoy M and Dresden M 1970 Statistical mechanics of the XY
model: I. {\it Phys. Rev. A} {\bf 2} 1075\nonum Baruch E and McCoy M
1971 Statistical mechanics of the XY model: II. Spin-correlation
functions {\it Phys. Rev. A} {\bf 3} 786\nonum Baruch E and McCoy M
1971 Statistical mechanics of the XY model: III. {\it Phys. Rev. A}
{\bf 3} 2137

\bibitem{jsp116bqj2004} Jin B-Q and Korepin V E 2004 Entanglement, Toeplitz determinants and Fisher-Hartwig
conjecture {\it J. Stat. Phys.} {\bf 116} 79

\bibitem{cmp252jpk2004} Keating J P and Mezzadri F 2004 Random-matrix theory and entanglement in quantum
spin chains {\it Commun. Math. Phys.} {\bf 252} 543\nonum Keating J
P and Mezzadri F 2005 Entanglement in quantum spin chains, symmetry
classes of random matrices, and conformal field theory {\it Phys.
Rev. Lett.} {\bf 94} 050501

\bibitem{jpa38ari2005} Its A R, Jin B-Q and Korepin V E 2005 Entanglement in the XY spin chain {\it J.
Phys. A} {\bf 38} 2975

\bibitem{preprint} Keating J P, Mezzadri F and Novaes M 2006 Comb
entanglement in quantum spin chains {\it Preprint} quant-ph/0604016

\bibitem{line} Kurmann J, Thomas H and M\"uller G 1982 Antiferromagnetic long-range order in the anisotropic
quantum spin chain {\it Physica A} {\bf 112} 235 \nonum M\"uller G
and Shrock R E 1985 Implications of direct-product ground states in
the one-dimensional quantum XYZ and XY spin chains {\it Phys. Rev.
B} {\bf 32} 5845


\end{thebibliography}
\end{document}